\documentclass[11pt]{epiarticle}

\usepackage{epimath}
\setpapertype{A4}
\usepackage{hyperref}
\usepackage[english,french]{babel}
\addto\captionsfrench{%
   \def\figurename{Figure}%
   \def\tablename{Tableau}%
}
\usepackage[T1]{fontenc}
\usepackage[utf8]{inputenc}
\usepackage{caption}

\newtheorem{definition}{Définition}

\usepackage{apacite}

\title{L'algorithme~: pourquoi et comment le définir pour l'enseigner}
\author{Emmanuel Beffara}
\date{}
\journal{\'EpiDEMES  -- (2024), 3} 
\acceptation{soumis le 02/06/2023, accepté le 18/10/2024}

\newcommand\siecle[1]{\textsc{#1}\up{e} siècle}
\begin{document}
\maketitle

\begin{prelims}
  \selectlanguage{english}

  \def\abstractname{Abstract}
  \abstract{
    The question of the definition of what is an algorithm is recurrent. It is
    found in teaching, at different levels and particularly in secondary
    education because of the recent evolutions in high school, with immediate
    consequences in higher education. It is found in mediation, with the
    different meanings that the word “algorithm” is charged with in the media
    space. It is also found in research, with issues in different branches of
    computer science, from foundations in computability and complexity to
    applications in big data. Beyond the issue of definition, it is the raison
    d'être of the notion of algorithm that should be questioned: what do we
    want to do with it and what is at stake? It is by trying to specify this
    that we can identify didactic elements that are likely to help teach the
    algorithm, in interaction with mathematics or not, and to different
    audiences.
  }

  \keywords{Algorithm; Computational thinking; Didactic transposition}

  \medskip

  \selectlanguage{french}
  \def\abstractname{Résumé}
  \abstract{
    La question de la définition de ce qu'est un algorithme est récurrente. Elle
    se trouve dans l'enseignement, à différents niveaux et singulièrement le
    secondaire du fait des évolutions récentes au collège et au lycée, avec des
    conséquences immédiates dans le supérieur. Elle se trouve dans la médiation,
    avec les différents sens dont le mot \og algorithme\fg\ est chargé dans l'espace
    médiatique. Elle se trouve aussi dans la recherche, avec des enjeux dans
    différentes branches de l'informatique, depuis les fondements en calculabilité
    et complexité jusqu'aux applications dans le traitement des données massives.
    Au delà du problème de la définition, c'est la raison d'être de la notion
    d'algorithme qu'il convient de questionner: que veut-on en faire et de quels
    enjeux est-elle le nom ? C'est en cherchant à préciser cela que l'on peut
    identifier les éléments didactiques susceptibles d'enseigner l'algorithme, en
    interaction avec les mathématiques ou pas et à différents publics.
  }

  \def\keywordsname{Mots-Clés}
  \keywords{Algorithme; Pensée informatique; Transposition didactique}

  \selectlanguage{french}
  \tableofcontents
\end{prelims}

\section{Introduction}

L'objectif de cet article est de s'interroger sur la bonne définition à donner
au mot \emph{algorithme} dans un contexte d'enseignement, dans différents
contextes.
Savoir définir ce dont on parle est une nécessité dès qu'il s'agit de
développer un discours scientifique, or l'introduction encore récente dans
l'enseignement général de la notion d'algorithme, parmi d'autres notions
fondamentales de la \emph{science} informatique, répond à une volonté de faire
évoluer l'enseignement scientifique pour y intégrer ces notions.

Donner une définition n'est pas difficile en soi et de nombreux ouvrages en
proposent.
Le problème se situe plutôt dans la variété des définitions disponibles, car
chaque contexte d'utilisation correspond à des intentions et des contraintes
différentes, au point que certaines formulations et certains usages peuvent
sembler contradictoires.
Dans un contexte d'enseignement, cette variété suscite de surcroît un enjeu de
cohérence: si des définitions différentes peuvent être nécessaires dans des
disciplines ou à des niveaux différents, l'enseignante ou la formatrice doit
savoir dissiper l'incohérence en ayant conscience de la transposition
didactique qui est faite dans ces différents contextes.
Ces questions se posent bien entendu dans toute situation où un savoir expert
fait l'objet d'une transposition, particulièrement à un public large et non
spécialiste.

Pour aborder ces questions dans le cas de l'algorithme, nous adoptons une
démarche de \og vigilance épistémologique\fg\ à visée didactique, au sens
d'\citeA{artigue-1990-epistemologie}, en analysant les différentes
manifestations de la notion d'algorithme dans la littérature scientifique et
les usages afin de dégager les raisons d'être des différentes approches et les
relations qu'elles entretiennent.
Nous commençons en section~\ref{sec:objet} par décrire l'émergence de la
notion d'algorithme comme objet d'étude, pour en évoquer les origines et la
situer dans la science actuelle.
Nous identifions dans la section~\ref{sec:definition} différents types de
définitions qui en sont proposées dans la littérature et l'usage, afin d'en
caractériser les différences essentielles.
Nous poursuivons en section~\ref{sec:enseignement} par une comparaison de la
façon dont se positionnent différents niveaux d'enseignement, de l'école
primaire à l'université (dans le système français).
Nous concluons en section~\ref{sec:conclusion} par une réflexion sur la raison
d'être des définitions de la notion d'algorithme.

\section{L'algorithme, de la pratique à l'objet d'étude}
\label{sec:objet}

On ne cherche pas ici à faire une histoire des algorithmes, mais plutôt à
distinguer différentes phases dans l'élaboration de cette notion.
Il n'est d'ailleurs pas évident que ces phases soient fidèles à la chronologie
historique, il s'agit plutôt de mettre en évidence, dans l'esprit de la
dialectique outil-objet \cite{douady-1986-jeux}, les différents statuts que
l'algorithme prend successivement dans l'esprit d'un apprenant ou d'une
communauté.
Pour une approche historique, on pourra se référer à des ouvrages comme celui
de \citeA{chabert-2010-histoire}.

\subsection{Procédures codifiées}

La première phase de construction de la notion concerne la systématisation des
procédures de calcul et la recherche de l'extension de leur domaine de
validité.
On peut citer quelques étapes emblématiques, dans une longue histoire des
mathématiques et du calcul~\cite{chabert-2010-histoire}:

\begin{itemize}
\item
  les méthodes de calcul posé, notamment pour l'addition et la multiplication,
  liées à chaque système de numération (méthodes promues par les
  \emph{algoristes} exploitant la numération de position, à l'opposé des
  \emph{abacistes} qui utilisaient les abaques);
\item
  l'approximation des racines carrées, déjà étudiée chez les babyloniens (d'où
  vient la méthode qui porte aujourd'hui le nom de Héron d'Alexandrie);
\item
  le calcul des diviseurs communs chez Euclide, qui est souvent la première
  méthode de calcul qui reçoit explicitement le nom d'algorithme dans
  l'enseignement actuel;
\item
  la résolution d'équations chez Al-Khwarizmi (dont le nom est notoirement à
  l'origine du mot \og algorithme\fg), qui fait intervenir des décisions
  dépendant des données d'entrée.
\end{itemize}

Ces quelques cas ne sont que des exemples parmi d'autres.
Les méthodes systématiques existent depuis aussi longtemps que les
mathématiques mais ne sont probablement pas perçues comme objets en
elles-mêmes, elles sont de l'ordre du savoir-faire.
On n'éprouve pas le besoin de définir ce que serait une méthode \emph{en
général}, même si on trouve chez Al-Khwarizmi l'idée d'étudier les méthodes
pour elles-mêmes.

Il se pose quand même, dès l'époque de ces exemples, les premières questions
de nature algorithmique: on s'interroge sur la faisabilité de certaines choses
par des méthodes systématiques.
C'est le cas notamment des trois grand problèmes géométriques de l'Antiquité:
la quadrature du cercle, la trisection de l'angle et la duplication du cube.
Dans les trois cas, on se demande s'il est possible de construire une certaine
grandeur avec les outils de la géométrie, autrement dit on cherche un
algorithme pour résoudre un problème donné au moyen d'opérations élémentaires
explicitement choisies.
L'impossibilité de résoudre ces trois problèmes à la règle et au compas ne sera
établie que bien plus tard, quand la notion de construction sera suffisamment
formalisée pour établir des résultats négatifs.

\subsection{Machines programmables}

Parallèlement à l'élaboration de méthodes de calcul en mathématiques,
l'histoire des techniques a connu le développement de machines pour réaliser
diverses tâches.
Les automates existent depuis l'Antiquité, ce sont des machines qui
agissent de façon plus ou moins autonome selon un plan établi, mais
c'est à partir du \siecle{xvii} que l'idée de \emph{programmation}
apparaît plus distinctement.

D'une part, il y a les machines à calculer, à commencer par la Pascaline de
Blaise Pascal, dont la première réalisation remonte à 1645, et qui apporte
l'idée de mécanisation du calcul, là où bouliers et abaques servaient
essentiellement à aider la mémoire tout en laissant à l'opérateur humain toute
la charge de la manipulation.
Les \emph{machines à différences} sont imaginées plus tard comme extensions de
ce principe pour produire des tables de valeurs de fonctions.

D'autre part, au \siecle{xviii}, on voit apparaître l'idée d'un objet
qui porte explicitement de l'\emph{information} qu'un dispositif
mécanique utilise pour déterminer son action: ce sont les métiers à
tisser semi-automatiques.
Le premier métier à rubans perforés est dû à Bouchon en 1725, suivi par le
métier Jacquard à cartes perforées en 1801.
Il y a alors l'idée de transcrire une procédure dans un langage artificiel et
explicitement défini, c'est un précurseur de la notion de \emph{programme} au
sens informatique: un \emph{objet} identifié, une information qui détermine
l'action d'une \emph{machine} (alors que l'opérateur concerné par les
procédures codifiées évoquées plus haut est humain).

La machine analytique de Babbage, conçue en 1834 mais jamais entièrement
réalisée, combine les deux aspects: la mécanique fait des calculs guidés par
des instructions indiquées sur des cartes perforées.
De plus, certaines instructions permettent le branchement conditionnel, donc
les structures de contrôle: il est possible de transcrire dans le langage de
la machine les structures comme \og si \dots\ alors \dots\ sinon \dots\fg\ et
\og tant que \dots\ faire \dots\fg, ce qui rend la machine \emph{universelle}
pour le calcul (au sens où elle peut théoriquement calculer tout ce qui
pourrait être calculé avec tout autre dispositif mécanique).
C'est pour la machine de Babbage que fut écrit ce qui est considéré comme le
premier véritable programme informatique, œuvre d'Ada Lovelace en 1843,
destiné à calculer les nombres de Bernoulli et reposant sur une itération
conditionnelle \cite{dufour-2019-ada}.

\subsection{Calculabilité}

Certains problèmes d'aspect algorithmique, comme les grands problèmes
géométriques de l'Antiquité, ou encore la résolution d'équations algébriques
par radicaux, ont été résolus négativement au moyen de différentes notions
d'invariants.
Au début du \siecle{xx}, Hilbert énonce deux problèmes fondamentaux qui sont
eux aussi de nature algorithmique:
\begin{itemize}
\item
  la recherche d'une méthode de résolution des équations diophantiennes,
  dixième problème de la liste formulée au Congrès international des
  mathématiciens de 1900 \cite{hilbert-1902-problemes};
\item
  le problème de la décision, formulé en 1928, qui consiste à établir une
  procédure permettant de déterminer si un énoncé donné est un théorème en
  logique du premier ordre \cite{hilbert-1959-grundzuge}.
\end{itemize}

La résolution négative de ces deux problèmes, par \citeA{goedel-1931-ueber}
pour le second et par \citeA{matiyasevich-1993-hilbert} pour le premier,
nécessite la formalisation de la notion de calcul.
Plusieurs \emph{modèles de calcul} sont proposés, notamment par
\citeA{church-1941-calculi} avec le $\lambda$-calcul,
\citeA{turing-1936-computable} avec la machine qui porte son nom (évoquée plus
précisément en section~\ref{sec:turing}), ou
\citeA{herbrand-1930-recherches} avec les fonctions récursives qui sont au
cœur des travaux de Gödel.
L'équivalence établie entre ces modèles fait émerger l'idée que la
calculabilité est en fait indépendante du modèle de calcul et que tout modèle
physiquement plausible définirait la même notion, affirmation que l'on appelle
habituellement \emph{thèse de Church-Turing}.

L'algorithme est alors un objet d'étude en soi et on essaie de
l'approcher par des définitions formelles, notamment pour établir des
résultats d'impossibilité.
Parmi ceux-ci, il y a l'identification de problèmes dits \emph{indécidables},
pour lesquels on montre qu'il ne peut pas exister d'algorithme: outre les deux
problèmes mentionnés ci-dessus, citons l'exemple classique du \emph{problème
de l'arrêt} (déterminer si l'exécution d'un programme donné, pour une entrée
donnée, finira par s'arrêter ou pas) ou encore le problème consistant à
déterminer si une expression donnée, formée avec les quatre opérations et les
fonctions usuelles de l'analyse, est égale à 0
\cite{richardson-1969-undecidable}.

Plus tard, la théorie de la complexité raffine cette notion en tentant
de classifier les problèmes calculatoires selon leur difficulté
intrinsèque, vue comme borne inférieure de l'efficacité des algorithmes
susceptibles de les résoudre \cite{perifel-2014-complexite}.

\subsection{L'algorithme au delà du calcul}

Ainsi, la calculabilité a formalisé la notion de procédure de calcul pour
résoudre des problèmes de décidabilité puis a permis le développement de
la théorie de la complexité, fondée au départ sur l'idée de calcul
séquentiel.
La sens même du mot \emph{calcul} s'en est trouvé étendu, de la manipulation
de nombres à l'idée plus large de manipulation de symboles.
Mais la technique ne s'arrête pas au calcul sur machine séquentielle et les
ordinateurs ont été développés bien au delà de cet objectif initial: les
machines sont devenues capables de faire plusieurs calculs en même temps
(parallélisme), de communiquer avec un utilisateur en cours d'exécution
(notion d'interface humain-machine), de communiquer de l'information avec
d'autres dispositifs (capteurs et actionneurs) et d'autres ordinateurs
(réseaux), sans évoquer l'hypothèse de machines changeant fondamentalement la
façon de calculer comme les processeurs quantiques.
Ces différents développements s'accompagnent de techniques de
programmation différentes qui dépassent aussi le cadre des seules
procédures de calcul des origines, et ces techniques nécessitent
d'enrichir le cadre théorique de l'algorithmique par de nouvelles
notions et de nouveaux outils.

De ce fait, l'étude des algorithmes comporte de nombreux enjeux dans la
recherche actuelle, dans des domaines très variés, des plus théoriques, comme
la logique et la complexité, aux plus concrets, comme l'analyse de données
pour la décision et la planification, en passant par la cryptologie, le calcul
formel ou numérique, le traitement automatique des langues, l'analyse et
évaluation de performances, et d'innombrables autres sujets.

Ce qui unifie ces différents développements, et rend légitime de les inclure
dans le domaine général des algorithmes, est le fait qu'il s'agit toujours
de traiter des données par la manipulation de leur représentation symbolique.
Pour autant, la notion de calcul, comprise même au sens large de manipulation
symbolique destinée à obtenir un résultat à partir d'une donnée d'entrée, ne
suffit plus à décrire la portée de ces algorithmes lorsque leur but n'est plus
de produire un résultat mais d'agir sur le monde.
Le domaine d'application du concept d'algorithme doit donc être considéré
comme plus large que le calcul.

\section{Une définition problématique}
\label{sec:definition}

Au vu de ces différents statuts de l'algorithme, qui acquiert progressivement
la qualité d'objet, on s'intéresse maintenant aux définitions possibles de ce
qu'est un algorithme, selon les contextes et les usages.
Le substantif \emph{algorithmique} désigne le domaine de savoir consacré à
l'élaboration, l'étude et l'analyse des algorithmes, sa définition découle
donc implicitement de celle de l'algorithme.

\subsection{Définitions formelles}
\label{sec:turing}

Les résultats formels nécessitent une définition mathématique, ce qui
mène à la notion de modèle de calcul.
Parmi les modèles les plus classiques, on trouve la machine de Turing, qui
formalise l'idée d'un opérateur qui effectue un calcul en appliquant des
règles précises et systématiques.
On pourra penser à une multiplication posée ou à la résolution d'un système
d'équations: suffisamment précisée, la procédure se réduit à lire et écrire
des symboles à des endroits particuliers et à prendre des décisions en
fonction de ce qui est écrit.

L'espace de travail est modélisé comme un ruban illimité constitué de cases
qui peuvent contenir des symboles dans un alphabet donné.
L'opérateur ne considère qu'une case à la fois et le programme de calcul
prescrit l'action à effectuer à chaque étape en fonction du contenu de la
case: écrire un symbole dans la case, se déplacer sur le ruban, poursuivre le
programme à une autre étape.
Cela peut être transformé en une définition formelle, par exemple de la forme
suivante
(inutile de s'attarder sur le détail, notre but n'est que de donner une
idée de ce à quoi ressemble une telle définition):
\begin{definition}
  Une machine de Turing est un sextuplet $(\Sigma,\Gamma,Q,i,f,\delta)$ où
  $\Sigma$ est un ensemble fini non vide (l'alphabet d'entrée et de sortie),
  $\Gamma$ est un sur-ensemble fini de $\Sigma$ (l'alphabet de travail)
  contenant un élément distingué $\emptyset$ (le symbole vide) non élément de
  $\Sigma$,
  $Q$ est un ensemble fini non vide (les états),
  $i$ et $f$ sont des éléments de $Q$ (l'état initial et l'état final) et
  $\delta$ est une fonction de $Q\times\Gamma$ dans
  $Q\times\Gamma\times\{-1,0,1\}$ (la table de transition).
\end{definition}
À cette définition devront s'ajouter celle d'une configuration (qui comprend
un état de la machine, un contenu de ruban et une position de la tête de
lecture sur le ruban) ainsi que la façon dont la machine calcule (les règles
de passage d'une configuration à la suivante).
De là, on pourra obtenir une définition mathématique de ce que calcule une
machine de Turing et donc étudier les propriétés de cette notion de calcul.

La machine de Turing est un objet formel sur lequel on peut fonder
toute la théorie de la calculabilité et de la complexité algorithmique, comme
le font bon nombre de textes de référence.
Pourtant, sa description formelle n'est pas totalement satisfaisante si on
veut lui donner le statut de définition générale de la notion d'algorithme:
\begin{itemize}
\item
  Il y a dans la définition ci-dessus des choix assez arbitraires, comme le fait
  d'identifier un unique état final $f$, les possibilités de
  déplacement de la tête (l'ensemble $\{-1,0,1\})$, le fait d'avoir le même
  alphabet $\Sigma$ en entrée et en sortie, etc.
  On trouve dans la littérature de nombreuses définitions qui font des choix
  différents.
\item
  Pour un objet correspondant à la définition, il y a des détails
  absolument pas fondamentaux, comme le choix de l'alphabet de travail,
  la dénomination des états, etc.
\item
  Au delà de ces détails, il y a des choix particuliers de modélisation:
  le calcul est purement séquentiel et déterministe, avec une donnée d'entrée
  et une donnée de sortie, représentées par des suites de symboles pris dans
  l'alphabet $\Sigma$.
\item
  Plus largement, il y inadéquation entre l'objet ainsi défini et
  l'usage de la notion d'algorithme: dans l'algorithme d'Euclide, on
  n'éprouve pas le besoin de préciser l'alphabet de travail et la
  représentation des nombres dans cet alphabet.
  Le fait d'écrire les nombres en base dix, en binaire ou même en chiffres
  romains ne fait pas partie de ce qui caractérise cet algorithme.
\end{itemize}
Le même phénomène se retrouvera quel que soit le modèle de calcul: il y
aura forcément des choix de modélisation et des conventions pour la
représentation des données.
Cela est inhérent à l'idée de définition formelle.

Notons au passage que si la machine de Turing fait figure de modèle de
référence pour la notion de calcul, c'est en particulier pour la simplicité de
sa définition (même si cela n'est pas forcément évident pour un lecteur qui ne
serait pas habitué à lire des définitions de ce genre).
On aurait aussi bien pu choisir comme modèle un langage de programmation
quelconque, mais la définition formelle aurait été largement plus complexe et
essentiellement impossible à manipuler mathématiquement (spécifier
complètement des langages de programmation pour en démontrer des propriétés
est un travail de recherche en soi et il nécessite l'utilisation des logiciels
pour manipuler les définitions et les démonstrations, tellement celles-ci sont
techniques).
De fait, il y a une tension entre la facilité d'écriture des programmes, effet
de la richesse du langage, et la facilité de raisonnement et de démonstration,
que favorise au contraire la minimalité du langage.
Ainsi, les modèles de calcul sont des langages de programmation simplifiés à
l'extrême pour ne conserver que le minimum permettant de calculer.

La \emph{thèse de Church-Turing} affirme que tous les modèles de calcul qui
correspondraient à des choses physiquement réalisables, du moins dans
leur principe, sont équivalents, au sens où ce qui est calculable par un
modèle le sera par les autres
(ce n'est pas un théorème, parce que sa formulation n'est pas assez
précise pour le permettre, mais cette équivalence a effectivement été
démontrée pour tous les modèles de calculs précis qui ont été définis).
Elle peut se raffiner pour exprimer que
le choix de représentation des données que l'on fait dans chaque usage
de chaque modèle n'a pas d'importance pour ce qui est des algorithmes
que l'on peut transcrire.

Pour ces raisons, toute définition formelle de la notion d'algorithme
apparaît comme réductrice parce qu'elle fixe nécessairement des détails et des
contraintes qui ne s'imposent pas dans l'absolu.
En somme, poser une définition formelle au sens des mathématiques fait passer
à côté du sujet puisqu'on en vient à définir ce qu'est un programme.
Il semble donc plus pertinent de chercher une définition qui
s'attache plutôt au concept, c'est-à-dire caractériser le propriétés
opératoires de la notion d'algorithme, ainsi que les techniques et modes de
raisonnement qui s'y appliquent \cite{vergnaud-1990-theorie}.

\subsection{Définitions conceptuelles}

Dans leur ouvrage \emph{Introduction à l'algorithmique}, qui est une des
références classiques au niveau universitaire,
\citeA{cormen-2004-introduction} proposent la définition suivante:
\begin{quote}
  Voici une définition informelle du terme algorithme: procédure de calcul
  bien définie qui prend en entrée une valeur, ou un ensemble de valeurs, et
  qui donne en sortie une valeur, ou un ensemble de valeurs. Un algorithme est
  donc une séquence d'étapes de calcul qui transforment l'entrée en sortie.
\end{quote}
Cette définition met en avant le fait d'éviter l'ambiguïté et le fait de
concerner des valeurs prises dans des ensembles mathématiquement bien définis.
Elle reste cependant très vague (volontairement) puisqu'elle repose sur l'idée
de procédure de calcul, elle-même non définie.

Dans son ouvrage de référence \emph{The Art of Computer Programming},
\citeA{knuth-1969-fundamental} ne définit pas non plus formellement ce qu'est
un algorithme, mais introduit la notion par un exemple (l'algorithme
d'Euclide, notamment parce qu'il est le premier à porter ce nom dans la
littérature mathématique) avant de préciser des critères qu'une procédure doit
respecter pour être qualifiée d'algorithme:
\begin{itemize}
\item
  finitude: l'exécution d'un algorithme doit toujours se terminer après un
  nombre fini d'étapes;
\item
  définition précise: chaque étape doit être définie précisément, les
  actions doivent être spécifiées rigoureusement et sans ambiguïté pour
  chaque cas;
\item
  entrées: des valeurs, prises dans un ensemble d'objets spécifié, qui
  sont données avant que l'exécution ne commence;
\item
  sorties: une ou plusieurs valeurs produites à l'issue de l'exécution et qui
  ont une relation spécifiée avec les entrées;
\item
  effectivité: les opérations à accomplir doivent être assez basiques
  pour pouvoir être en principe réalisées en un temps fini par une
  personne utilisant un papier et un crayon.
\end{itemize}
Knuth fonde sur une machine idéalisée (MIX dans les premières éditions, MMIX
dans les éditions plus récentes) les notions nécessitant un modèle de calcul
précis, néanmoins la différence entre ce modèle et la notion d'algorithme est
ainsi explicitement posée.
Il en ressort donc le côté opérationnel (effectivité et non-ambiguïté) et, à
nouveau, le fait de concerner des valeurs prises dans des ensembles
mathématiquement bien définis.
La \og relation\fg\ spécifiée entre entrée et sortie est le plus souvent
une fonction au sens mathématique (il y a un unique résultat attendu pour
chaque donnée d'entrée) mais ce n'est pas une nécessité.

La définition proposée par \citeA{modeste-2012-enseigner} dans sa thèse
synthétise ces points, sur la base de différentes définitions dont celles
évoquées plus haut, en ajoutant explicitement la référence à un problème
que l'algorithme est censé résoudre:
\begin{quote}
  Un algorithme est une procédure de résolution de problème, s'appliquant
  à une famille d'instances du problème et produisant, en un nombre fini
  d'étapes constructives, effectives, non-ambiguës et organisées, la
  réponse au problème pour toute instance de cette famille.
  \cite[p. 25]{modeste-2012-enseigner}
\end{quote}
La notion de problème ici est à comprendre au sens informatique: c'est la
spécification d'un ensemble de valeurs d'entrées possibles (les instances du
problème) et d'une réponse attendue pour chaque entrée.
L'algorithme est alors vu comme une méthode permettant de calculer
effectivement cette réponse, sachant que la définition du problème permet de
caractériser la réponse attendue mais pas la façon de l'obtenir.
Cette mise en avant des notions de problème et d'instance souligne
l'importance de la \emph{généricité} attendue d'un algorithme, puisqu'il doit
traiter de façon systématique un ensemble généralement infini de cas, et
en corollaire la notion de \emph{domaine} de définition attendu (que l'on
pourrait aussi appeler sa portée), élément important dès que l'on cherche à
caractériser et comparer des algorithmes.

Le chapeau de l'article \emph{Algorithme} de l'encyclopédie en ligne
Wikipédia\footnote{\url{https://fr.wikipedia.org/wiki/Algorithme}, consulté le
21 novembre 2024}
donne une formulation concise cohérente avec celle de Modeste:
\begin{quote}
  Un algorithme est une suite finie et non ambiguë d'instructions et
  d'opérations permettant de résoudre une classe de problèmes.
\end{quote}
L'expression \og classe de problèmes\fg\ exprime ici l'idée de généricité,
comme le mot \og problème\fg\ au sens plus précis évoqué plus haut,
d'une façon seulement plus explicite pour être accessible à un
lectorat moins expert.
Des définitions similaires à celle-ci sont données par les dictionnaires
généralistes (notamment le Larousse, le Robert, le TLFi\footnote{Le TLFi
  mentionne aussi le sens archaïque de \og Système de numération décimale en
  chiffres arabes\fg\ avec les règles opératoires qui s'y rapportent, ce que
l'on retrouve avec le mot \emph{algoriste} évoqué plus haut.}
ainsi que le Wiktionnaire, chacun consultés en juin 2024).
D'autres variantes, notamment issues de documents destinés aux enseignants,
reprennent ce principe d'une procédure destinée à résoudre un problème.
Ainsi, dans leur brochure destinée aux enseignants de mathématiques en classe
de seconde, \citeA{sopena-2012-initiation} écrivent:
\begin{quote}
  Un algorithme décrit un enchaînement d'opérations permettant, en un temps
  fini, de résoudre toutes les instances d'un problème donné. Un algorithme
  permet donc, à partir d'une instance du problème (les données en entrée),
  d'obtenir un résultat correspondant à la solution du problème sur cette
  instance. Ce résultat est obtenu en réalisant \og pas à pas\fg\ une
  succession d'opérations élémentaires.
  \cite[p. 7]{sopena-2012-initiation}
\end{quote}
Cette formulation met en avant l'idée d'opération \og élémentaire\fg, ce qui
reprend l'idée d'effectivité dans les critères de \citeA{knuth-1969-fundamental}.

Cette comparaison de différentes définitions informelles révèle donc une
différence fondamentale avec les définitions formelles évoquées plus haut: la
référence à la notion de \emph{problème}.
Dans la définition de Knuth (\emph{ibid.}), la \og relation spécifiée\fg\ entre les entrées
et les sorties, bien que plus allusive, dit quelque chose de similaire: une
suite d'opérations prend le statut d'algorithme quand on précise la nature des
entrées et des sorties ainsi que la façon dont la sortie est censée dépendre
de l'entrée.
Ainsi, selon ces définitions, le texte suivant est bien une procédure
effective et non ambiguë mais pas un algorithme, car il ne contient pas de
\emph{spécification}:
\begin{tabbing}
  \qquad\= Entrées: $a$ et $b$, entiers naturels \\
  \>Tant que $b\neq 0$, \\
  \>\qquad si $a>b$ \= alors retrancher $b$ à $a$ \\
  \>\> sinon échanger $a$ et $b$. \\
  \>Renvoyer $a$.
\end{tabbing}
Ce texte devient un algorithme si sa spécification est explicitée, en l'espèce
si on exprime que la valeur renvoyée doit être le plus grand diviseur commun
de $a$ et $b$, ou si on annonce qu'il est censé résoudre le problème du calcul
du PGCD.

Le fait que la spécification fasse partie de l'algorithme ne signifie pas que
la procédure sans spécification ne soit pas un objet digne d'intérêt.
C'est même plutôt le contraire: une part conséquente de l'activité
algorithmique consiste justement à étudier des procédures pour déterminer si
elles respectent certaines spécifications, c'est la question de la correction
des algorithmes.
Ainsi, le fait d'inclure la spécification dans la définition du mot
\emph{algorithme}, ce qui peut être discuté (et ne fait probablement pas
consensus), correspond surtout au besoin de distinguer la procédure
\emph{brute} et l'intention qu'on lui attribue, un élément au cœur des
compétences liées à l'algorithmique (voir section~\ref{sec:competences}).

Selon ces définitions, il y a donc deux dimensions importantes: d'une part le
fait qu'il y ait une procédure, plan d'action ayant une certaine généricité et
décrite de façon finie, et d'autre part le fait qu'il y ait un problème à
résoudre, donc une finalité.
La finitude de la description est cruciale et impose de distinguer clairement
la procédure de son exécution: dès que le domaine d'entrée est infini, il est
impossible d'identifier un algorithme à l'ensemble de ses exécutions, la façon
de décrire la procédure est donc partie intégrante de ce qui caractérise un
algorithme.
Certaines définitions laissent ce point assez implicite (comme celle de
Wikipédia citée plus haut qui évoque une \og suite [...] d'instructions et
d'opérations\fg) alors que la dualité entre l'objet statique (la description)
et le comportement dynamique (l'exécution sur une entrée) est fondamentale.

\subsection{L'usage dans les médias}

Au-delà des tentatives de définitions, qu'elles soient formelles ou
informelles, il est indispensable d'évoquer l'usage médiatique du mot \og
algorithme\fg, étant donnée la place croissante qu'il a pris ces dernières
années.
Comme dans le cas de beaucoup de termes issus du vocabulaire
scientifique repris par les médias, le sens attribué à ce mot est très
flou, probablement autant pour les rédacteurs que pour les lecteurs.
On peut néanmoins identifier quelques traits significatifs.

Le mot est principalement utilisé pour désigner des systèmes informatiques
utilisés dans de nombreux contextes de le vie publique, de l'administration à
la finance, et notamment les systèmes de captation et de monétisation de
l'attention développés dans les réseaux sociaux et sites de commerce,
ou encore les systèmes de surveillance.
On pourra citer quelques titres glanés dans la presse parmi de nombreux
autres: \og Consulter Facebook sans algorithme et sans contenu toxique, le
pari fou du Digital Services Act\fg\ (\emph{Le Soir}, 25 août 2023), \og
Comment le complotisme prospère grâce à l'algorithme des réseaux sociaux\fg\
(\emph{Ouest France}, 27 mai 2023), \og Contre le terrorisme, le fantasme de
l'algorithme préventif\fg\ (\emph{Le Monde}, 12 mai 2021), \og La ``ville
sûre'' ou la gouvernance par les algorithmes\fg\ (\emph{Le Monde
diplomatique}, juin 2019).
Il est clair que le mot \og algorithme\fg\ s'écarte ici des définitions
précédentes, non seulement parce qu'il est difficile d'identifier précisément
un problème à résoudre (au sens de la section précédente), mais aussi parce
que l'intérêt n'est pas porté sur les méthodes systématiques de traitement de
données mais sur le phénomène d'automatisation et ses effets.

Par ailleurs, il y a souvent une confusion entre \og algorithme\fg\ et \og
intelligence artificielle\fg.
Cette expression, dont on connaît le potentiel médiatique, est elle-même prise
dans un sens plutôt restreint qui correspond aux dernières applications
connues du grand public, notamment l'apprentissage profond, le traitement des
grandes masses de données ou les applications des grands modèles de langage.
Il est inutile de préciser qu'une telle interprétation ne rend justice ni à
l'algorithmique ni à l'intelligence artificielle, domaines dont la richesse et
la diversité excèdent largement ces quelques thèmes.

Il est tout à fait normal que ces sujets préoccupent les médias et
l'ensemble des citoyens, puisque ces systèmes affectent le quotidien de
tout le monde avec des effets préoccupants dont le grand public prend
progressivement conscience; les effets des bulles informationnelles sur
les comportements de consommation et sur les processus démocratiques
sont d'ailleurs bien connus et documentés~\cite{ertzscheid-2017-appetit}.
C'est un manque de culture scientifique qui mène à cette confusion dans le
vocabulaire médiatique, quand on désigne du nom savant d'algorithme un objet
fantasmé, mystérieux et inquiétant, sorte de cerveau immatériel qui prendrait
des décisions sur la vie de chacun de façon autonome et incontrôlée.

Ce ne sont pas \og les algorithmes\fg\ qui nous gouvernent.
Les infrastructures informatiques des réseaux sociaux sont conçues par des
humains qui y impriment leur vision du monde et ce sont des humains qui font
le choix de les utiliser dans le fonctionnement de leurs plateformes.
Savoir correctement définir la notion d'algorithme est indispensable pour
pouvoir prendre le recul nécessaire pour comprendre ces enjeux.

\section{L'algorithme comme objet d'enseignement}
\label{sec:enseignement}

La transposition didactique \cite{chevallard-1991-transposition} est le
phénomène selon lequel un savoir \emph{savant}, notion définie et utilisée par
les experts d'un domaine, est transformé pour en faire un savoir
\emph{enseigné}, notion adaptée au contexte didactique, c'est-à-dire aux
connaissances et savoir-faire supposés de la part du public ciblé et aux
objectifs d'apprentissage choisis.
Ce phénomène inhérent à l'enseignement est le résultat d'une action plus ou
moins consciente et volontaire guidée par la contrainte que la notion à
enseigner doit être accessible tout gardant une certaine fidélité à la notion
savante.
Par ailleurs, cette fidélité est en partie une \og fiction\fg\
\cite{artigue-1990-epistemologie} parce que les savoirs savants et enseignés
obéissent à des dynamiques différentes qui peuvent justifier certaines
infidélités.

Ainsi, dans le cas de l'algorithme, des approches assez différentes se
rencontrent selon les niveaux.
Au delà des contraintes propres de la transposition didactique, la variété
d'approches correspond à différentes intentions que l'on peut mettre dans
l'enseignement de la notion, c'est-à-dire différentes raisons qui poussent à
l'inclure dans les disciplines d'enseignement et leurs programmes.
Dans cette partie, on s'intéresse à aux approches que l'on rencontre dans
l'enseignement en France.


\subsection{Dans l'enseignement supérieur}

Dans l'enseignement supérieur, la notion d'algorithme se trouve dans les
filières pour lesquelles l'informatique fait partie des objectifs
d'apprentissage, comme discipline principale ou secondaire.
On y vise donc un savoir expert, qu'il s'agit de construire avec très peu
d'effet de transposition.
Selon que les filières concernées sont plus orientées vers la pratique ou vers
la théorie, il s'agit d'apprendre à de futures informaticiennes ou
scientifiques à concevoir leurs programmes ou bien de poser les bases pour
comprendre des résultats de calculabilité et de complexité.
On retrouvera donc les algorithmes classiques et les techniques pour les
programmer, la démonstration de correction des algorithmes par rapport à leur
spécification et l'analyse de la complexité des algorithmes et des problèmes.

Même si on n'éprouve pas le besoin de donner une définition précise de ce
qu'est un algorithme, comme en témoignent les choix de
\citeA{knuth-1969-fundamental} ou de \citeA{cormen-2004-introduction}, on voit
en fait l'algorithme comme l'idée derrière la conception d'un programme.
En d'autres termes, l'objectif est l'implémentation sous forme d'un programme
sur une machine, l'algorithme est la description de haut niveau de
l'organisation d'un programme, indépendamment des aspects liés spécifiquement
à la mise en œuvre (détails de syntaxe du langage utilisé, choix des
structures de données à employer, utilisation de bibliothèques logicielles).
La programmation sert donc de justification et de référence et les résultats
formels portant sur la correction et la complexité sont fondés sur l'intuition
partagée par les étudiants après plusieurs années de pratique de la
programmation.
Les modèles de calcul définis mathématiquement n'apparaissent dans les cursus
universitaires qu'en fin de licence ou en master, en s'appuyant sur cette
pratique de la programmation, dans le but de fonder les notions de
calculabilité et de complexité des problèmes.

Une étude de \citeA{rafalska-2018-conception} sur les préconceptions des
élèves et étudiants quant à la notion d'algorithme montre que, à partir de la
licence, l'idée d'algorithme est fondamentalement liée à la notion de problème
à résoudre: on n'envisage pas ce que serait une procédure de calcul
indépendamment de ce qu'elle est censée calculer.
De fait, à partir de la deuxième année de licence, la notion de preuve
d'algorithme, c'est-à-dire de démonstration de l'adéquation entre la procédure
décrite et une spécification donnée indépendamment, prend une importance
majeure; on étudie un algorithme pour démontrer qu'il est correct.
À cela s'ajoute le critère d'efficacité dans la qualité d'un algorithme, qui
est présent dans les conceptions des étudiants dès la première année de
licence.

\subsection{Dans l'enseignement secondaire}

Dans l'enseignement obligatoire en France (jusqu'à 16 ans),
l'algorithmique est explicitement présente depuis les programmes de 2016.
Que ce soit dans l'enseignement primaire ou secondaire, de tronc commun ou
optionnel, elle est abordée comme un objet de culture générale scientifique,
contribuant à développer la \og pensée informatique\footnote{J'emploie ici le
  mot \emph{informatique} comme traduction de l'anglais \emph{computational}
  faute de mieux: on rencontre parfois le mot \emph{computationnel} dans des
  textes en français mais c'est plus une adaptation phonétique de l'anglais
  qu'une traduction, quant au mot \emph{calculatoire} qui en est une
  traduction correcte dans certain cas, il est trop restrictif dans le
  contexte qui nous intéresse.}\fg\ telle que la présente
  \citeA{wing-2006-computational}.
Néanmoins, le rapport à l'algorithme change fondamentalement entre le cycle~3
(transition du primaire au collège) et le cycle~4 (trois dernières années du
collège) et les définitions pertinentes doivent être adaptées aux attentes.

On se concentrera donc dans cette section sur le cycle~4, où la notion
d'algorithme est installée pour l'usage qui en est fait dans l'enseignement de
tronc commun jusqu'à la seconde générale.
On pourra se reporter à la brochure de la CII Lycée
\cite{beffara-2017-algorithmique} pour une étude plus détaillée des notions en
jeu à ce niveau.
On ne se penchera pas explicitement sur le cas du lycée parce que la
notion y est traitée en continuité avec le collège, sauf pour certains sujets
dans les disciplines de spécialité touchant à l'informatique où ce sont les
approches suivies dans le supérieur qui seront pertinentes (par exemple pour
la sensibilisation à la preuve de correction ou la décidabilité, qui font
partie des programmes de la spécialité \og Numérique et sciences
informatiques\fg).

Au collège, la notion d'algorithme est introduite explicitement et elle se
construit en même temps que la pratique de la programmation, comme en témoigne
le \og Thème E -- Algorithmique et programmation\fg\ du programme de
mathématiques de cycle~4, dont les attendus de fin de cycle indiquent:
\begin{quotation}
  \emph{Écrire, mettre au point, exécuter un programme}

  Connaissances
  \begin{itemize}
    \item Notions d'algorithme et de programme.
    \item Notion de variable informatique.
    \item Déclenchement d'une action par un événement.
    \item Séquences d'instructions, boucles, instructions conditionnelles.
  \end{itemize}
  Compétences associées
  \begin{itemize}
    \item Écrire, mettre au point (tester, corriger) et exécuter un programme
      en réponse à un problème donné.
  \end{itemize}
  \cite[annexe 3]{menj-2020-bo-c234}
\end{quotation}
Dans l'esprit des documents institutionnels pour le collège, les notions
d'algorithme et de programme sont implicitement distinguées, mais l'algorithme
est essentiellement vu comme un brouillon de programme: sa raison d'être est
l'implémentation qui en sera faite.
De plus, le logiciel Scratch est utilisé systématiquement comme référence (il
n'est pas mentionné explicitement dans les programmes officiels mais son
influence y est visible, de plus il est explicitement utilisé dans les
documents d'accompagnement et dans les sujets du brevet).

Les choix des différents manuels pour le cycle~4 quant à la définition de la
notion d'algorithme sont variés, mais des traits communs se retrouvent.
La distinction entre algorithme et programme est toujours présente, même si
elle est plus ou moins claire selon les documents, et parfois formulée de
façon discutable.
Ainsi, le \emph{Cahier d'algorithmique et de programmation Cycle 4} (éd.
Delagrave, 2016, pp.~4-5), écrit:
\begin{quote}
  Je comprends
  \begin{itemize}
    \item Un \emph{algorithme} est une suite d'\emph{instructions} à appliquer
      dans un ordre logique pour résoudre un problème et obtenir rapidement%
\footnote{%
On considèrera la référence à la rapidité dans cette définition comme une
simple maladresse: l'intention est certainement de suggérer que l'efficacité
des algorithmes est une préoccupation importante, mais il est bien clair
qu'une méthode systématique et non ambiguë qui permettrait d'obtenir un
résultat sans que ce soit rapide mériterait tout autant d'être appelée
algorithme.}
      un résultat. Il est écrit à la main ou à l'aide d'un logiciel dans un
      langage compréhensible par tous.
  \end{itemize}
  Je retiens
  \begin{itemize}
    \item Un algorithme sert à préparer l'écriture d'un programme
      informatique.
  \end{itemize}
\end{quote}
De façon similaire, le manuel \emph{Transmath cycle 4} (éd. Nathan, 2016,
p.~548) écrit:
\begin{quote}
  Je retiens
  \begin{itemize}
    \item Un algorithme décrit la démarche logique d'un programme. Il met en
      évidence la structure de ce programme et fait apparaître ses variables.
    \item Un fois mis au point, l'algorithme est codé dans un langage de
      programmation.
  \end{itemize}
\end{quote}
On comprend de ces formulations que l'algorithme est considéré comme une
méthode, par opposition au programme qui est une mise en œuvre concrète, et
que le fait qu'il soit communicable est mis en avant.

En revanche, ces deux définitions suggèrent une méthode de travail qui
consiste en deux phases bien marquées et distinctes: premièrement élaborer un
algorithme sur papier jusqu'à ce qu'il soit parfaitement au point,
deuxièmement le programmer sur ordinateur.
Cette vision est discutable tant du point de vue épistémologique que du point
de vue pratique.
D'une part, l'écriture d'un programme n'est pas la seule raison d'être d'un
algorithme: l'algorithme de l'addition posée est étudié et pratiqué en détail
par les élèves à l'école primaire, mais on ne cherche pas à le faire
programmer aux élèves.
D'autre part, une informaticienne qui élabore un algorithme ne s'interdira pas
d'en programmer des ébauches dans un but d'expérimentation, car c'est ce qui
permet d'acquérir des intuitions sur son fonctionnement.
Néanmoins, identifier l'élaboration de l'algorithme et sa programmation comme
deux activités différentes est pertinent et utile pour l'enseignement car ces
activités mettent en œuvre des compétences différentes.

Quelques manuels évitent la référence à la programmation comme un passage
obligé et emploient des analogies avec des situations où l'ordinateur
n'intervient pas.
Ainsi, le manuel \emph{Myriade Cycle 4} (éd. Bordas, 2016, p.~20) propose:
\begin{quote}
  Un \emph{algorithme} est une suite finie d'instructions permettant de
  résoudre un problème.

  Exemple: Une recette de cuisine est un algorithme.

  En effet, on dispose d'ingrédients au départ, on applique les instructions
  données par la recette et on obtient le plat désiré à la fin.
\end{quote}
Si l'image avec la recette de cuisine est tentante et permet de faire
comprendre l'idée de suite d'opérations communicable, il convient de prendre
garde à ne pas pousser trop loin l'analogie, pour éviter de créer des
confusions (les ingrédients ne sont pas des \emph{données} d'entrée, le
problème à résoudre est difficile à identifier, etc.).

Au delà de la question de définir la notion d'algorithme, les manuels
scolaires du secondaire se focalisent sur les structures perçues comme
fondamentales: les structures de contrôle (séquence, condition, itération), la
notion de variable, et plus tard les structures de données (listes et
tableaux, au lycée) et la notion de fonction au sens informatique (avec les
notions d'abstraction, décomposition et généralisation qu'elle porte).
L'enjeu de la correction des algorithmes est évoqué dans les dernières années
de l'enseignement secondaire sans être un sujet central: ce sont la
compréhension des techniques et le savoir-faire algorithmique qui sont les
objectifs principaux.

\subsection{Dans l'enseignement primaire}

À l'école élémentaire, les sujets explicitement informatiques sont plutôt
anecdotiques mais les prémices de l'algorithmique sont présents.
La notion d'algorithme n'est explicitement mentionnée en mathématiques
qu'en référence aux procédures du calcul posé, mettant donc en avant l'enjeu
de développer un savoir-faire calculatoire.
Ainsi, le programme officiel indique:
\begin{quote}
  \emph{Calcul posé.}
  Connaître et mettre en œuvre un algorithme de calcul posé pour effectuer :
  \begin{itemize}
    \item l'addition, la soustraction et la multiplication de nombres entiers
      ou décimaux ;
    \item la division euclidienne d'un entier par un entier ;
    \item la division d'un nombre décimal (entier ou non) par un nombre
      entier.
  \end{itemize}
  \cite[annexe 2]{menj-2020-bo-c234}
\end{quote}
La notion est également vue comme objet à découvrir dans le
cadre des apprentissages en \emph{Sciences et technologie}:
\begin{quote}
  Les élèves découvrent l'algorithme en utilisant des logiciels d'applications
  visuelles et ludiques.
\end{quote}

Néanmoins, des activités de programmation sont évoquées, en lien avec le
repérage.
Ainsi, dans la partie \emph{Espace et géométrie} du programme de mathématiques
de cycle~2, on trouve parmi les attendus détaillés de fin de cycle:
\begin{quote}
  (Se) repérer et (se) déplacer en utilisant des repères et des
  représentations

  [\dots]
  \begin{itemize}
    \item Programmer les déplacements d’un robot ou ceux d’un personnage sur
      un écran:
      \begin{itemize}
        \item repères spatiaux;
        \item relations entre l'espace dans lequel on se déplace et ses
          représentations.
      \end{itemize}
  \end{itemize}
  \cite[annexe 1]{menj-2020-bo-c234}
\end{quote}
L'esprit de cet attendu, qui est confirmé dans les diverses ressources
proposées aux enseignants comme dans les pratiques en classe, est de mettre en
avant la notion de suite d'opérations et les deux premières compétences
fondamentales liées à l'informatique que sont l'anticipation et l'évaluation
(voir section~\ref{sec:competences}).
En effet, à la différence de l'attendu sur les algorithmes de calcul posé où
il n'est question que de connaître et savoir appliquer un algorithme donné,
les problèmes de programmation de déplacement mettent en jeu la planification
d'un trajet et sa formulation au moyen de codes et de conventions.
Ceux-ci
peuvent être fournis, comme dans le cas du langage de commande d'un robot ou
des instructions élémentaires de Scratch, ou élaborés par l'élève, comme dans
les activités telles que le \emph{robot idiot} \cite{romero-2018-jeu}.

À l'école maternelle, le mot \og algorithme\fg\ est employé par les programmes
\cite{menj-2020-bo-c1} pour désigner ce qu'on désignerait dans un langage plus
courant comme des suites logiques, parfois qualifiées dans ce contexte de \og
suites algorithmiques\fg.
Dans ce contexte, l'initiation à une pensée informatique ne fait pas partie des
objectifs (à juste titre, car distinguer une forme de pensée typiquement
informatique à l'âge de l'école maternelle n'est certainement pas pertinent).
Néanmoins, l'attention est portée sur l'identification de régularités dans un
processus, en lien avec l'idée de rythme (travaillée aussi avec le corps, par
la musique et la percussion) et de répétition.
L'attendu qui consiste à savoir identifier un motif et le reproduire peut
être envisagé comme un précurseur de l'identification de \og motifs\fg\
opératoires, plus tard formalisés en algorithmes.
L'utilisation de suites logiques, par opposition aux situations de déplacement
pratiquées à l'école élémentaire, a l'avantage de ne pas recourir au repérage
spatial, compétence qui est en cours de construction à cet âge et se développe
encore dans les années suivantes \cite{leonard-2021-motif}.

\subsection{Les compétences de l'algorithmique}
\label{sec:competences}

Comme on l'a vu, les façons d'aborder l'objet algorithme sont très variées
d'un niveau d'enseignement à l'autre, ce qui pose la question de la cohérence
vis-à-vis de la notion sous-jacente.
L'approche didactique, en analysant les compétences en jeu dans les tâches
considérées comme liées à l'algorithmique, permet de dégager de grands
principes qui permettent de faire le lien entre les différentes approches.
On reprend ici la structuration en cinq compétences élaborée notamment par
Declercq à l'occasion de la mise en place des formations d'enseignants lors de
l'introduction de la nouvelle discipline \og Numérique et sciences
informatiques\fg\ au lycée général \cite{declercq-2021-didactique}.

\emph{Anticiper} consiste à se mettre en posture de programmeur pour décrire
l'enchaînement d'une suite d'opérations, avant le début de l'exécution.
C'est un élément fondamental du passage du statut d'outil à celui d'objet pour
l'algorithme, puisqu'il s'agit, pour reprendre l'analyse de
\citeA{samurcay-1985-faire}, de passer de \emph{faire} une suite d'opérations
pour résoudre un problème à \emph{faire faire} ces opérations à un autre
opérateur (humain ou mécanique) à un autre moment.
Cette planification, déconnexion entre le temps de l'écriture et celui de
l'exécution, est un trait fondamental de l'élaboration d'algorithmes.

\emph{Évaluer} consiste attribuer mentalement une valeur à un programme donné.
Le mot \og valeur\fg\ est à prendre dans un sens très large: il peut s'agir
bien entendu de suivre un algorithme pour en obtenir le résultat, mais aussi
d'évaluer le nombre d'opérations à effectuer en fonction de l'entrée, de
déterminer si le résultat respectera une spécification donnée, etc.
Là encore, il s'agit de considérer l'algorithme comme un objet, et en
particulier de le considérer pour ce qu'il est (avec ses erreurs éventuelles)
plutôt que pour ce qu'on voudrait qu'il soit.
Se dégager de l'intentionnalité est l'enjeu des activités d'initiation comme
le \emph{robot idiot} autant que des problèmes de démonstration de correction
qui se rencontrent aux niveaux avancés.

\emph{Décomposer} est une compétence liée à la résolution de problème, elle
consiste à transformer un problème complexe en un ensemble de problèmes plus
simples dont l'agencement permet de résoudre le problème initial.
Cette compétence se retrouve de façon très voisine dans l'enseignement des
mathématiques, sa spécificité dans le cas algorithmique est peut-être de
raisonner en termes de décomposition dans le temps.

\emph{Généraliser} consiste à inférer un schéma général à partir d'une ou
plusieurs instances particulières.
Cela inclut la reconnaissance de motifs, enjeu de certaines activités d'école
primaire et de collège notamment, mais aussi la capacité à repérer dans un
problème particulier la répétition de traitements ou de données suivant un
même schéma.
Par la suite, la capacité à généraliser consiste à formuler un problème
général à partir d'une instance, pour en rendre le traitement plus
systématique ou le rapprocher de schémas connus (par exemple en voyant le
calcul du 100ème terme de la suite Fibonacci comme un cas particulier du
calcul du $n$-ième terme d'une récurrence linéaire d'ordre 2).

\emph{Abstraire} consiste à sélectionner l'information utile pour la
résolution d'un problème donné, et donc à \og faire abstraction\fg\ des
informations non pertinentes.
Cela met donc en jeu la notion d'information, caractéristique de la science
informatique (qui en tire son nom), puisqu'un algorithme traite uniquement de
l'information, sous forme de données bien définies.
À un niveau avancé, l'abstraction est ce qui permet de créer des solutions où
la manière de résoudre un problème peut être \og abstraite\fg\ à l'aide d'une
interface pertinente (par exemple, rechercher une valeur dans une liste de
nombres croissante ou un mot dans un dictionnaire peuvent être vus comme deux
instances d'une recherche dans une liste triée, l'interface étant cette de la
structure d'ordre).

Bien entendu, la plupart des tâches algorithmiques impliquent plusieurs de ces
compétences, qu'il s'agisse d'élaborer ou d'analyser des algorithmes.
C'est l'association de ces différentes dimensions qui caractérise l'activité
propre à l'algorithmique et permet en fin de compte d'établir une cohérence
entre les enseignements à différents niveaux scolaires et répondant à
différents objectifs.

\section{Conclusion: pourquoi définir}
\label{sec:conclusion}

Comme on l'a vu, on peut tenter de définir l'algorithme de différentes façons,
plus ou moins formelles.
La variété de définitions n'exclut pas la cohérence et elle correspond à
différents objectifs.
En fin de compte, il convient de se demander ce que l'on veut tirer d'une
définition.
Pour cette raison, il ne semble pas pertinent de conclure cette réflexion par
une définition unique qui en ferait la synthèse.

Pour obtenir des résultats mathématiques de calculabilité, comme des théorèmes
d'indécidabilité ou des bornes de complexité, il faut un modèle de calcul. 
C'est un élément de savoir savant qui sert de fondement aux autres approches,
comme la logique formelle sert de fondement à la preuve mathématique: c'est un
point de référence théorique mais le domaine ne s'y réduit pas.
Si l'on cherche un moyen de spécifier, valider et analyser des méthodes de
calcul, il faut plutôt caractériser un type de discours univoque et
opératoire.
Un tel discours pourrait \emph{en principe} être précisé jusqu'à être
implémenté dans un modèle de calcul (qu'il s'agisse d'un modèle théorique ou
d'un langage de programmation concret), mais cela ne nécessite pas la
définition précise d'un tel modèle.
Là encore, c'est un élément de savoir savant qui suppose un fondement
théorique précis et un environnement scientifique qui permet de le faire
vivre.
Si la définition est envisagée comme un élément de culture scientifique sur
les méthodes de calcul et la conception de programmes, il faut alors mettre en
avant le rôle des notions connexes de donnée, information, problème, machine
(ce qui permet ensuite d'aborder la notion savante en en comprenant les
enjeux).
Il s'agit dans ce contexte d'introduire à la pensée informatique
\cite{wing-2006-computational} en s'appuyant sur les grandes notions qui
fondent la science informatique \cite{dowek-2011-quatre}.
On a alors un élément de savoir à enseigner qui est le résultat d'une
transposition didactique de la notion savante, dans un contexte institutionnel
particulier.
Si l'objectif est de transmettre un élément de culture générale pour
contribuer à démystifier l'informatique, les définitions opératoires
évoquées précédemment sont secondaires, l'enjeu important est de suggérer le
genre de questions qui se rapportent à la notion d'algorithme et les façons de
penser qui y sont associées.
On est dans ce cas dans une forme de médiation qui donne à apercevoir un
domaine de connaissance et justifie son apprentissage.

Au delà de ces différentes façons d'aborder la notion d'algorithme, 
il convient bien entendu de se poser la question de la pertinence de cet
enseignement.
Si la recherche d'une \og raison d'être\fg\ de l'enseignement de
l'algorithme, pour reprendre les termes de \citeA{chiprianov-2018-enseigner},
n'est pas l'objet de nos réflexions ici, c'est une question inextricablement
liée à celle de la définition de l'objet et de l'orientation à donner à son
enseignement.
S'il est admis que travailler l'algorithme, au même titre que tout autre objet
de savoir scientifique, peut contribuer à apprendre à raisonner avec rigueur, 
il est possible que connaître de l'algorithmique n'ait que peu d'effet sur
la capacité à utiliser des dispositifs numériques avec aisance et esprit
critique.
Quelle que soit l'opinion que l'on peut avoir sur l'intérêt d'enseigner
l'informatique à tous les élèves et sur l'âge à partir duquel ce serait
pertinent, la présence grandissante du mot \og algorithme\fg\ dans le discours
public et la sphère médiatique l'impose comme un élément de culture générale
sur lequel il est indispensable que chaque citoyen ait quelques notions justes.

\bibliographystyle{apacite}
\bibliography{demimes}

\authoraddresses{
Emmanuel Beffara\\
Univ. Grenoble Alpes, CNRS, Grenoble INP, LIG\\
38000 Grenoble, France\\
\email emmanuel.beffara@univ-grenoble-alpes.fr
}

\end{document}